# Suppression of Hydrodynamic Escape of an $H_2$-rich Early Earth Atmosphere by Radiative Cooling of Carbon Oxides

Tatsuya Yoshida[1]

Corresponding author

Email: tatsuya@tohoku.ac.jp

Naoki Terada[1]

Email: teradan@tohoku.ac.jp

Kiyoshi Kuramoto[2]

Email: keikei@ep.sci.hokudai.ac.jp

(Institutional addresses)

[1] Graduate School of Science, Tohoku University, Sendai, Miyagi 980-8578, Japan

[2] Faculty of Science, Hokkaido University, Sapporo, Hokkaido 060-0810, Japan

# Abstract

Radiative cooling by molecules is a crucial process for hydrodynamic escape, as it can efficiently remove the thermal energy driving the outflow, acquired through X-ray and extreme UV absorption. Carbon oxides, such as CO and $CO_2$, and their photochemical products are anticipated to serve as vital radiative cooling sources not only in atmospheres dominated by carbon oxides but also in $H_2$-rich atmospheres. However, their specific effects on the hydrodynamic escape especially in $H_2$-rich atmospheres have been inadequately investigated. In this study, we conduct 1-D hydrodynamic escape simulations for $H_2$-rich atmospheres incorporating CO, $CO_2$, and their chemical products on an Earth-mass planet. We consider detailed radiative cooling processes and chemical networks related to carbon oxides to elucidate their impacts on the hydrodynamic escape. In the escape outflow, $CO_2$ undergoes rapid photolysis, producing CO and atomic oxygen, while CO exhibits photochemical stability compared to $CO_2$. The $H_2$ oxidation by atomic oxygen results in the production of OH and $H_2O$. Consequently, the hydrodynamic escape is significantly suppressed by the radiative cooling effects of CO, $H_2O$, OH, and $H_3^+$ even when the basal mixing fraction of CO and $CO_2$ is lower than ~0.01. These mechanisms extend the lifetime of $H_2$-rich atmospheres by about one order of magnitude compared to the case of pure hydrogen atmospheres on early Earth, which also results in negligible escape of heavier carbon- and nitrogen-bearing molecules and noble gases.

# 1 Introduction

Hydrodynamic escape is an important process of atmospheric loss that can drastically change atmospheric compositions (e.g., Lammer et al., 2008; Tian, 2015; Owen, 2019; Gronoff et al., 2020). Hydrogen-rich atmospheres exposed to high stellar X-ray and extreme ultraviolet (XUV) radiation are particularly susceptible to hydrodynamic escape. On Earth, an H$_2$-rich proto-atmosphere may have formed by impact degassing associated with simultaneous accretion of reduced materials (e.g., Kuramoto and Matsui, 1996; Schaefer and Fegley, 2010) and/or gravitational capture of solar nebular gas (e.g., Nakazawa et al., 1985; Ikoma and Genda, 2006). Earth may have lost abundant H$_2$ through hydrodynamic escape induced by strong XUV irradiation from the young Sun to change the atmospheric composition enriched in heavier molecules (e.g., Sekiya et al., 1980; Lammer et al., 2014; Yoshida and Kuramoto, 2021). Beyond our solar system, evidence for H$_2$-dominated atmospheres on exoplanets with mass as low as ~2 Earth mass has been accumulated (e.g., Owen et al., 2020) and the occurrence of hydrodynamic escape has been indicated for gaseous exoplanets close to their parent stars (e.g., Vidal-Madjar et al., 2003; Vidal-Madjar et al., 2004).

The escape rate of $H_2$ is a key factor controlling the lifetime and compositional evolution of an $H_2$-rich atmosphere. In addition to the XUV heating, another important process for regulating the hydrodynamic escape rate is the radiative cooling of expanding gas, which can remove the thermal energy gained by XUV absorption. Among various coolants, C-bearing molecules are considered important radiative cooling sources due to their optically active nature in infrared wavelengths. In reduced proto-atmospheres on Earth, $CH_4$ and its photochemical products such as CH and $CH_3$ can significantly reduce the $H_2$ escape rate under which the escape of C-bearing species becomes negligible (Yoshida and Kuramoto, 2021), where $CH_4$ has strong bands near 3.3 μm and 7.7 μm, CH has a key band near 4.0 μm, and $CH_3$ has strong bands near 3.3 μm and 15 μm, respectively (Rothman et al. 2013; Tennyson et al. 2016). As well as hydrocarbons, carbon oxides CO and $CO_2$ have high thermal emission rates over a wide range of the typical temperature of escape outflows (Figure 1). Here CO has a strong fundamental band near 4.7 μm, and $CO_2$ has key bands near 4.3 μm and 15 μm (Rothman et al. 2013; Tennyson et al. 2016). Yoshida and Kuramoto (2020) showed that the radiative cooling by CO, as well as $CH_4$, can significantly suppress the hydrodynamic escape of a Martian $H_2$-rich proto-atmosphere. In $CO_2$-rich atmospheres on terrestrial planets, 15 μm emission by $CO_2$ has been estimated to contribute to a decrease in the thermospheric temperature and escape rate (Kulikov

et al., 2006, 2007; Tian et al., 2008, 2009; Johnstone, 2018; Zahnle et al., 2019; Johnstone et al., 2021; Nakayama et al., 2022). This would be the case also for $H_2$-rich atmospheres contaminated by $CO_2$.

Carbon oxides can be supplied to an $H_2$-rich atmosphere on proto-Earth through $CH_4$ oxidation by $H_2O$-photolysis-derived oxidant radicals (Zahnle et al., 2020; Wogan et al., 2023; Yoshida et al., 2024), volcanic degassing from the oxidized upper mantle (e.g., Armstrong et al., 2019; Deng et al., 2020; Kuwahara et al., 2023), and impact degassing of small bodies containing carbonates (e.g., Yokoyama et al., 2022) and carbon oxide ices (Mumma and Charnley, 2011). On planets like Earth heavier than Mars, photolysis of C-bearing species should proceed efficiently in their deeper gravitational well (Yoshida and Kuramoto, 2021). In this case, radiatively active photochemical products like $H_3^+$, OH, and $H_2O$ derived from carbon oxides may also serve as important radiative cooling sources.

In this study, we update our 1-D hydrodynamic escape model by considering detailed radiative cooling processes and chemical networks related to carbon oxides to clarify their effects on the hydrodynamic escape of $H_2$-dominated atmospheres. The rest of this paper is organized as follows. In Section 2, we describe the outline of our hydrodynamic escape model. In Section 3, we show the numerical results of the atmospheric profiles, energy balance, and atmospheric escape rate. In Sections 4.1 and 4.2, we discuss the dependence of the atmospheric escape rate on the XUV flux and the planetary mass, respectively. In Section 4.3, we compare the results of $H_2$-CO and $H_2$-$CO_2$ atmospheres with those of $H_2$-$CH_4$ atmospheres. In Section 4.4, we discuss the evolution of early Earth's atmosphere based on the calculation results of this study.

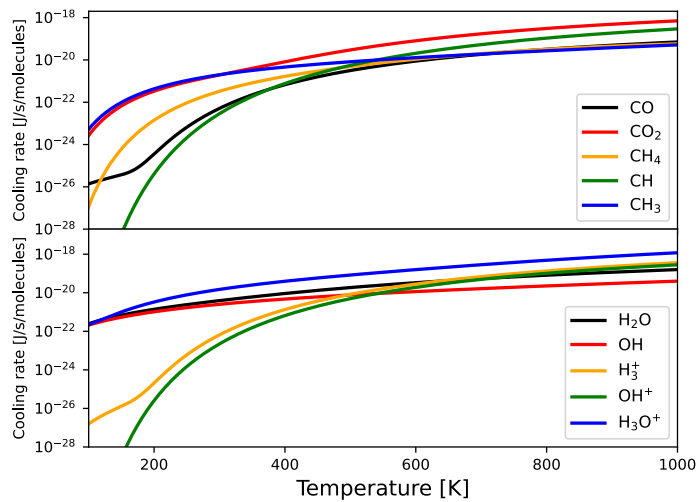

Figure 1. Radiative cooling rate per molecule under optically thin and LTE conditions as a function of temperature. The upper panel shows the cooling rates of C-bearing species, while the lower panel displays those of H- and O-bearing species, respectively. The net radiative cooling rate of each species is calculated by summing the cooling rate for each energy transition (Equation (12) in Supplementary Information), with reference to line data provided by HITRAN and ExoMol database.

## 2 Model description

A radially 1D hydrodynamic escape model developed by Yoshida and Kuramoto (2020) is applied with some updates to chemical and radiative processes. The details of the model are described in Supplementary Information. Below, we describe the outline of our model.

We suppose $H_2$-rich atmospheres including CO, $CO_2$, and their chemical products on a planet with the Earth's mass and radius. The realistic atmosphere may contain other background species including hydrocarbons, nitrogen-bearing species, $H_2O$ leaked from the lower atmosphere, and so on, but we simplify the compositional setup to clarify the effects of carbon oxides. The lower boundary of the upper atmosphere, which corresponds to the homopause, is set at $r = R_p+1000$ km, where $r$ is the radial distance from the planetary center and $R_p$ is the planetary radius, supposing the formation of geometrically thick $H_2$-rich atmospheres as assumed by Yoshida and Kuramoto (2021). The number density of $H_2$ at the lower boundary is set at $1 \times 10^{19}$ m$^{-3}$ supposing the typical gas density at the altitude above which the stellar XUV irradiation is absorbed completely (e.g., Kasting and Pollack 1983). The basal number densities of CO and $CO_2$ are given as parameters in each simulation run. The temperature at the lower boundary is set at 200 K approximating the atmospheric skin temperature of Earth under the faint young Sun. The upper boundary is set at $r = 30R_p$. The other physical quantities such as gas velocities at the lower and upper boundaries are estimated by linear extrapolations from the inside of the calculation domain.

The basic equations of this model are the fluid equations of continuity, momentum, and energy for a multi-component gas considering chemical processes and radiative processes on the assumption of spherical symmetry (Supplementary Information). They are solved by numerical integration about time until the physical quantities settle into steady profiles by the same calculation method as Yoshida and Kuramoto (2020) (see also Supplementary Information).

To describe chemical processes, 287 chemical reactions are considered for 30 atmospheric species: $H_2$, $CH_4$, $CO$, $CO_2$, $CH$, $CH_2$, $CH_3$, $H_2O$, $OH$, $O_2$, $H$, $C$, $O$, $O(^1D)$, $H^+$, $H_2^+$, $H_3^+$, $C^+$, $CH^+$, $CH_2^+$, $CH_3^+$, $CH_4^+$, $CH_5^+$, $O^+$, $CO^+$, $CO_2^+$, $O_2^+$, $OH^+$, $H_2O^+$, and $H_3O^+$. Here we add O-bearing species and their chemical reactions, building on the chemical network established by Yoshida and Kuramoto (2021) by referring to McElroy et al. (2013) and Yoshida et al. (2022). We neglect the formation of molecules that have more than one carbon because of low atmospheric densities in the atmospheric region where outflow to space accelerates. See Supplementary Information for the model formulation of chemical processes.

This study considers radiative heating through solar X-ray and UV absorptions by adopting the solar spectrum from 0.1 to 280 nm at the age of 100 Myr estimated by Claire et al. (2012). The total XUV flux from 0.1 nm to 100 nm is about 0.58 W/m$^2$ at the orbit of Earth, which is about 100 times as large as the present flux from the quiescent Sun. To calculate the profiles of heating and photolysis rates, we numerically solve the radiative transfer of parallel stellar photon beams in the spherically symmetric atmosphere by applying the method formulated by Tian et al. (2005). Following this calculation, the spherical shell average is taken for the 3D heating distribution. We neglect stellar visible and IR absorption as a heating source since stellar X-ray and UV absorption are expected to be the primary drivers of heating in the outflow.

We consider radiative cooling by thermal line emission of CO, $CO_2$, $H_2O$, OH, $CH_4$, CH, $CH_3$, $H_3^+$, $OH^+$, and $H_3O^+$ with line data provided by HITRAN database (Rothman et al. 2013; http://hitran.org) and ExoMol database (Tennyson et al. 2016; http://exomol.com). Here we consider 16061 transitions of $CO_2$ and use the same selected lines for CO as Yoshida and Kuramoto (2020), for $CH_4$, CH, and $CH_3$ as Yoshida and Kuramoto (2021), for $H_2O$ and OH as Yoshida et al. (2022) to cover more than 99% of the total energy emission in the temperature range of 100–1000 K. The line emission by $H_3^+$, $OH^+$, and $H_3O^+$ are assumed to be optically thin in the whole calculation region, considering all the lines in the ExoMol database. Although $CH_2$ would be also an important infrared active molecule, this study does not include its contribution to the radiative cooling because there are little $CH_2$ line data at present. We calculate the radiative cooling rate of the neutral radiative species by applying the method formulated by Yoshida and Kuramoto (2020). As discussed in Yoshida and Kuramoto (2020) and Ito et al. (2024), molecular emissions dominate in the lower thermosphere, where collisional transitions tend to occur efficiently. Therefore, we assume local thermal equilibrium (LTE) conditions to estimate the radiative cooling rate, although non-LTE effects can reduce the efficiency of radiative cooling, particularly in low-pressure regions (e.g., García Muñoz et al., 2024). See Supplementary Information for the model formulation of radiative transfer.

## 3 Results

The steady-state profiles of radial velocity, temperature, number density, and heating/cooling rate on $H_2$-CO atmospheres and $H_2$-$CO_2$ atmospheres are shown in Figures 2 and 3. Here we define $H_2$-CO atmospheres and $H_2$-$CO_2$ atmospheres as atmospheres that include a given fraction of CO and $CO_2$ at the lower boundary, respectively. In each simulation run, the atmospheric flow is accelerated from near zero to supersonic (Figures 2(a) and (b)) through the heating by absorption of solar X-ray and UV irradiation (Figures 3(e) and (f)).

The behaviors of CO and $CO_2$ in the outflow are different from each other. $CO_2$ is more efficiently photolyzed to produce CO and atomic oxygen (Figure 3(b)) due to its strong UV absorption with a wide range of wavelengths shorter than ~230 nm. On the other hand, CO is photochemically stable compared with $CO_2$, as it absorbs only UV with a wavelength shorter than ~100 nm. Therefore, CO becomes the main C-bearing species in the lower region with $r <$ ~$2R_p$ even in the $H_2$-$CO_2$ atmospheres (Figures 3(a) and (b)).

The production of atomic oxygen through photolysis of CO and $CO_2$ leads to the production of important coolants $H_2O$ and OH (Figures 3(a) and (b)) mainly by the following

reactions:

$$H_2 + O \rightarrow H + OH \quad (1)$$

$$H_2 + OH \rightarrow H_2O + H \quad (2)$$

$$OH + OH \rightarrow H_2O + O \quad (3)$$

The radiative cooling by various coolants works efficiently in both $H_2$-CO atmospheres and $H_2$-$CO_2$ atmospheres. The major coolants in the region $r < \sim 2R_p$ are $H_2O$ and CO while $H_3^+$ becomes the main coolant in the upper region (Figures 3(e) and (f)). Due to the radiative cooling effects, the temperature of the outflow becomes lower on the whole as the basal $CO/H_2$ or $CO_2/H_2$ ratio increases (Figures 2(c) and (d)).

To clarify the energy balance in the whole outflow, the heating/cooling rate and heating efficiency in the subsonic region are shown in Figure 4. Here the heating efficiency, the fraction of absorbed radiative energy partitioned to net atmospheric heating, is given by

$$\eta = \frac{\int_{r_0}^{r_s}(q_{\text{abs}} - q_{\text{ch}} - q_{\text{rad}})4\pi r^2 dr}{\int_{r_0}^{r_s} q_{\text{abs}} 4\pi r^2 dr} \quad (4)$$

where $r_0$ and $r_s$ are the radial distance of the lower boundary and the transonic point, respectively, and $q_{\text{abs}}$, $q_{\text{ch}}$, and $q_{\text{rad}}$ are the radiative heating rate by X-ray and UV absorption, the rate of the net chemical energy expense, and the radiative cooling rate. The heating efficiency decreases as the basal $CO/H_2$ or $CO_2/H_2$ ratio increases due simply to the increase in the total radiative cooling rate with the abundance of coolants other than $H_3^+$. The major coolants in the whole subsonic region are $H_3^+$ and CO. The radiative cooling rate of $CO_2$ is low compared with the major coolants due to its efficient photolysis. $H_2O$ and OH also enhance the energy loss by radiative cooling, especially in the $H_2$-$CO_2$ atmospheres where the production of atomic oxygen by $CO_2$ photolysis occurs efficiently. With increasing CO/$H_2$ or $CO_2$/$H_2$ ratio, the heating efficiency decreases and becomes almost constant when CO/$H_2$ or $CO_2$/$H_2$ > ~0.001. The latter is because the abundances of CO, $CO_2$, and their chemical products present in the outflow approach a limit controlled by the critical flux of escaping hydrogen as described below.

The atmospheric escape rate decreases rapidly with heating efficiency as the basal mixing ratio of CO or $CO_2$ increases, particularly when the ratio is lower than ~0.001 as shown in Figure 5, due to the enhanced radiative cooling by CO and photochemical products such as OH and $H_2O$. Here the bulk mass escape rate $F_{m,i}$ and molecular escape rate $F_i$ are related as follows:

$$F_{m,i} = m_i F_i = m_i 4\pi r_0^2 n_i(r_0) u_i(r_0), \qquad (5)$$

where $m_i$, $n_i(r_0)$, and $u_i(r_0)$ are the molecular mass, number density, and velocity of species *i*. Note that the fluxes of photochemical products are negligible at the lower boundary $r = r_0$ although most molecules are lost as dissociated atoms at the upper boundary.

On the other hand, when the basal mixing ratio of CO or $CO_2$ exceeds ~0.001, the escape rate becomes nearly constant and asymptotes to the critical flux, which is the minimum flux of H$_2$ required to drag other heavier species outside the planetary gravitational well and equivalent to the diffusion-limited escape flux (Figure 5). The critical flux for species $i$ is given by

$$F_{\text{crit},i} = \frac{4\pi r_0^2 b_{\text{H}_2,i} g X_{\text{H}_2}}{k_B T}(m_i - m_{\text{H}_2}) \qquad (6)$$

where $F_{\text{crit},i}$ is the critical flux for species $i$, $b_{\text{H}_2,i}$ is the binary diffusion coefficient between H$_2$ and species $i$, $g$ is the gravitational acceleration, $X_{\text{H}_2}$ is the mixing fraction of H$_2$, $k_B$ is the Boltzmann constant, and $T$ is the atmospheric temperature (Hunten et al., 1987). The critical flux is based on the crossover mass, which is the largest molecular mass of secondary species $i$ that can be dragged upward by the dominant lighter gas:

$$m_{c,i} = m_{\text{H}_2} + \frac{k_B T}{b_{\text{H}_2,i} g X_{\text{H}_2}} \frac{F_{\text{H}_2}}{4\pi r_0^2}. \qquad (7)$$

Equation (6) is derived by setting $m_{c,i} = m_i$ in Equation (7). Note that both the crossover mass and critical flux assume an isothermal atmosphere composed of only two species with constant vertical composition (Hunten et al., 1987). The convergence to the critical flux is because the H$_2$ escape flux hardly becomes lower than the critical flux unless the escape flow leaves carbon oxides behind at low altitudes and efficiently accelerates at high altitudes in weak radiative cooling. Under the critical flux, C-bearing species have a radial mixing profile adjusted by the balance between the upward drag force by escaping hydrogen and the gravitational force (Hunten et al., 1987). The heating efficiency also converges with increasing the basal $CO/H_2$ or $CO_2/H_2$ ratio so that the $H_2$ escape flux approaches the critical flux. The escape rate of $H_2$ when the basal $CO_2/H_2 > \sim 0.001$ is slightly lower than the critical flux because $CO_2$ is rapidly photolyzed in the outflow while the critical flux is derived neglecting $CO_2$ decomposition.

As a result of the decrease in the escape rate with increasing the basal mixing ratio of CO or $CO_2$, the fractionation factor between $H_2$ and the carbon oxides, defined as $(F_i/F_{H_2})/(n_i(r_0)/n_{H_2}(r_0))$ (where $i =$ CO, $CO_2$), also decreases as shown in Figure 6. This indicates that carbon oxides tend to be retained while $H_2$ escapes, especially when the $H_2$ escape flux approaches the critical flux. The dashed lines in Figure 6 represent fractionation factors derived from the analytical formula expressed by the crossover mass (Hunten et al., 1987):

$$\frac{F_i/F_{H_2}}{n_i(r_0)/n_{H_2}(r_0)} = \frac{m_{c,i} - m_i}{m_{c,i} - m_{H_2}}. \qquad (8)$$

The numerical results align closely with the semi-analytical values from Equation (8) for $H_2$-CO atmospheres and at low $CO_2/H_2$ ratios in $H_2$-$CO_2$ atmospheres. The deviation from the analytical formula observed at higher $CO_2/H_2$ ratios may be due to molecular decomposition, which violates the assumption of a two-component atmosphere with a constant vertical mixing ratio required by Equation (8), potentially allowing the $H_2$ escape flux to fall below the critical flux.

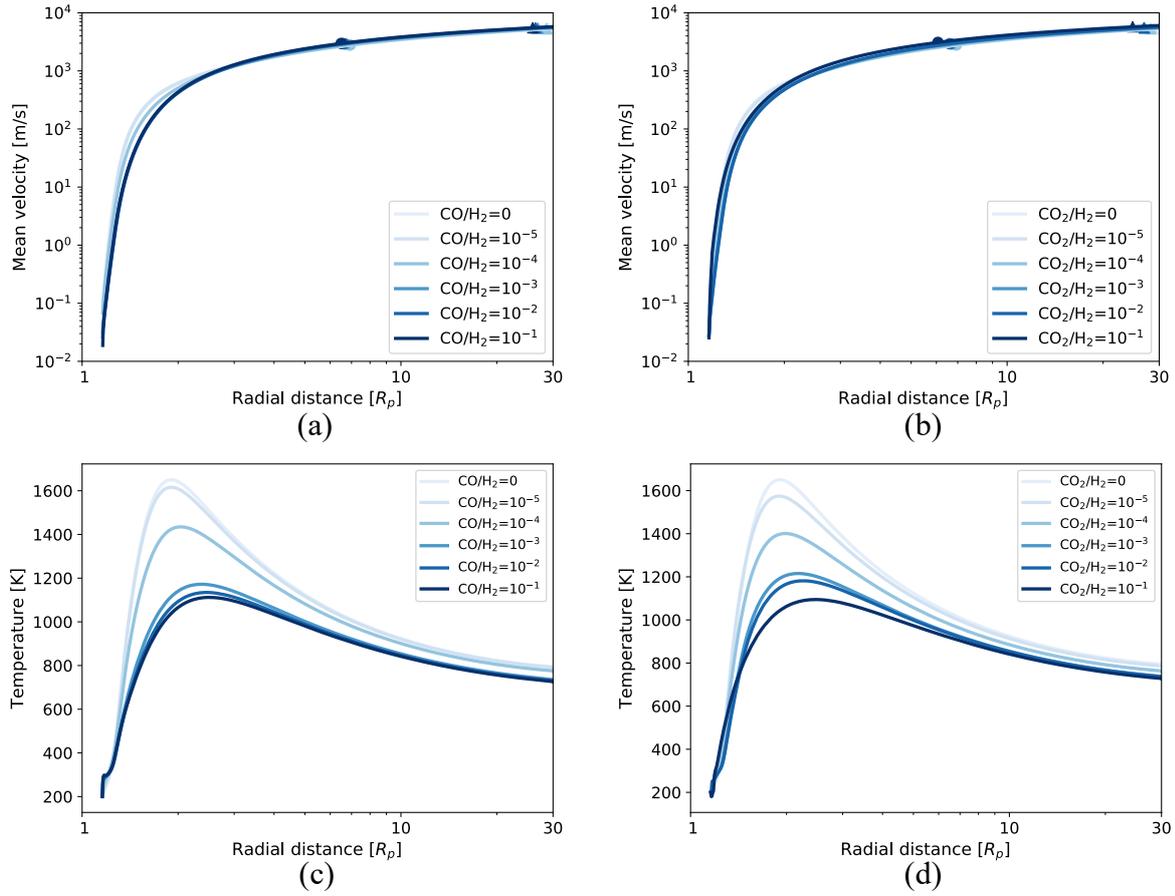

Figure 2. Velocity and temperature profiles of the $H_2$-CO atmospheres (left column) and $H_2$-$CO_2$ atmospheres (right column). (a): Mean velocity of the $H_2$-CO atmospheres. The dots and triangles represent the transonic points and exobases, respectively. (b): Mean velocity of the $H_2$-$CO_2$ atmospheres. (c): Temperature of the $H_2$-CO atmospheres. (d): Temperature of the $H_2$-$CO_2$ atmospheres.

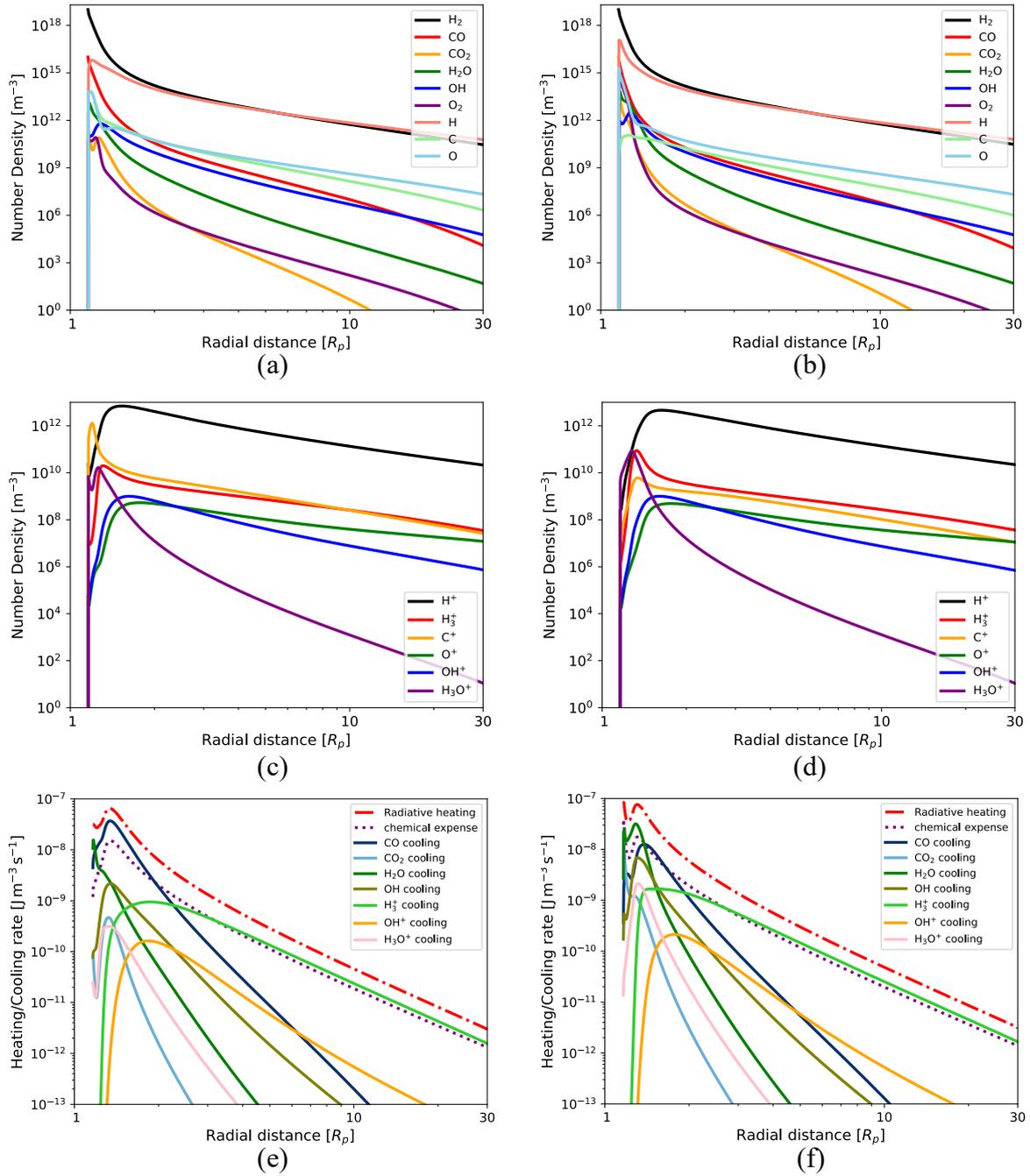

Figure 3. Number density and heating/cooling rate profiles of the $H_2$-CO atmospheres with the basal CO/$H_2$=0.001 (left column) and $H_2$-$CO_2$ atmospheres with the basal $CO_2$/$H_2$=0.001 (right column). (a): Number densities of selected neutral species in the $H_2$-CO atmosphere. (c): Number densities of selected ion species in the $H_2$-CO atmosphere. (e): Radiative heating rate (red dash-dotted line), radiative cooling rates (solid lines), and chemical energy expense rate (purple dotted line) in the $H_2$-CO atmosphere, respectively. (b), (d), and (f) are the same as (a), (c), and (e), respectively, but for the $H_2$-$CO_2$ atmospheres.

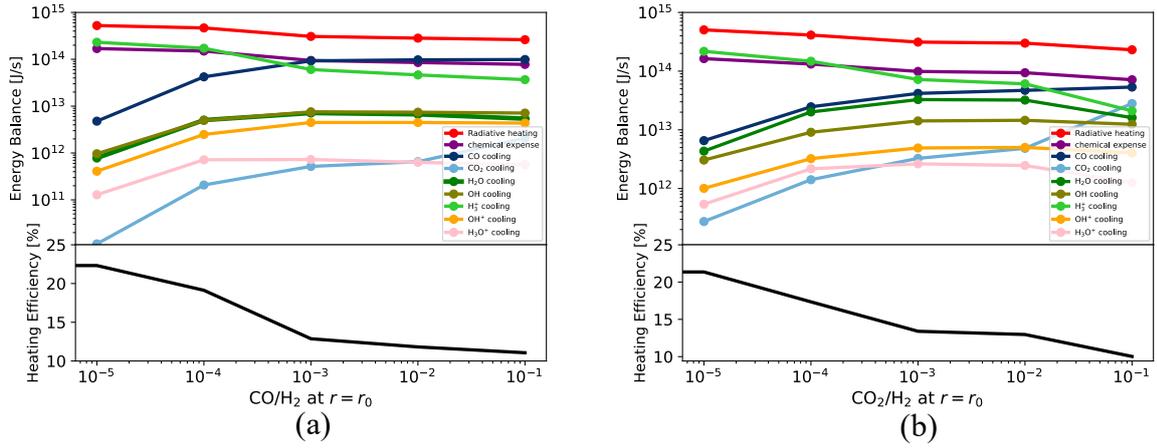

Figure 4. (a) Rates of radiative heating, radiative cooling, and chemical energy expense (upper panels) and heating efficiency defined as Equation (4) (lower panels) in the subsonic region of the $H_2$-CO atmospheres as functions of the basal CO/$H_2$ ratio. (b) Same as (a), but for the $H_2$-$CO_2$ atmospheres.

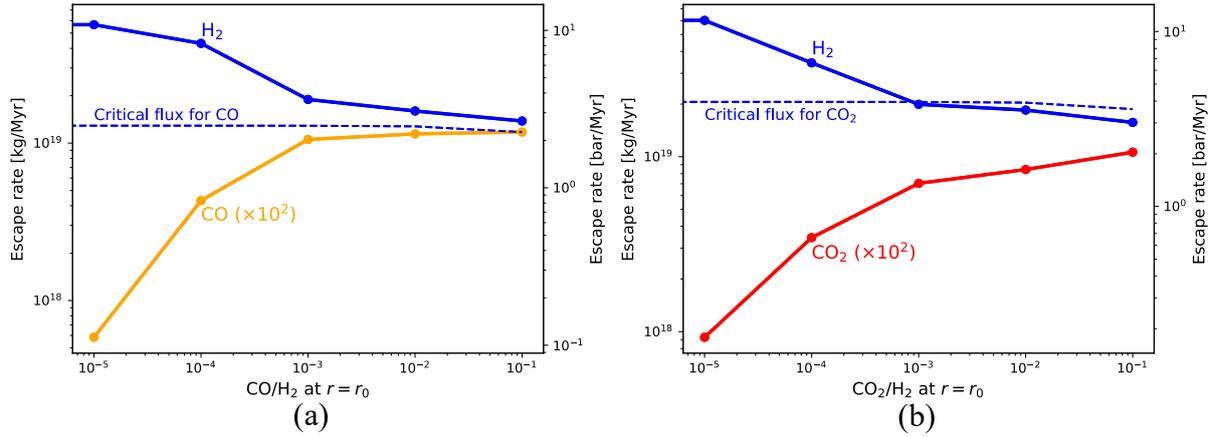

Figure 5. (a) Escape rate of main species per 1 Myr as a function of the basal CO/$H_2$ ratio on the $H_2$-CO atmospheres. (b) Same as (a), but for the $H_2$-$CO_2$ atmospheres. The escape rates of CO (orange) and $CO_2$ (red) are multiplied by 100. The right vertical axis represents the mass flux normalized by the mass of the present-day Earth's atmosphere: 1 bar = $5.3 \times 10^{18}$ kg. The dashed lines show the critical flux, which are the minimum fluxes of $H_2$ that can drag other heavier species outside the planetary gravitational well, as described by Equation (6).

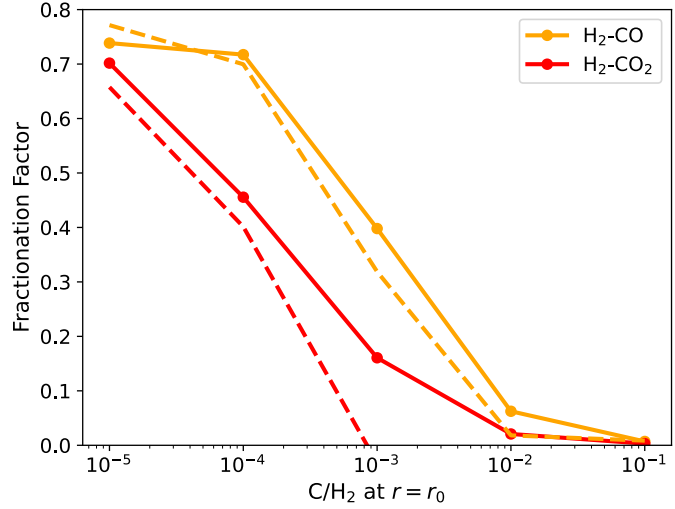

Figure 6. Fractionation factor, defined as $(F_i/F_{H_2})/(n_i(r_0)/n_{H_2}(r_0))$ (where $i = $ CO, $CO_2$), as a function of the basal $CO/H_2$ or $CO_2/H_2$ ratio. Solid lines represent the numerical results, while dashed lines show the values derived from the analytical formula in Equation (8). The orange lines represent the fractionation factors between $H_2$ and CO in $H_2$-CO atmospheres, and the red lines represent those between $H_2$ and $CO_2$ in $H_2$-$CO_2$ atmospheres.

# 4 Discussion

## 4.1 Effects of a change in XUV flux

In the previous section, we adopt the X-ray and UV spectrum estimated for young Sun with the XUV flux of about 100 times the present-day value. However, the estimates of the XUV flux for the young Sun have significant uncertainties (Tu et al., 2015). Moreover, the XUV flux would have been highly variable as observed in exoplanet-hosting young stars (e.g., Johnstone et al., 2021). This section investigates the effects of a change in XUV flux on the atmospheric escape rate.

Figure 7 compares the atmospheric escape rates under conditions where the XUV flux is twice the standard case with those under the standard XUV setting. The escape rate increases almost proportionally with the XUV flux when the basal $CO/H_2$ or $CO_2/H_2$ ratio is < ~0.001. On the other hand, the escape rate eventually asymptotes the critical flux even under the higher XUV flux due to the energy loss by radiative cooling. Thus, radiative cooling can also suppress hydrodynamic escape under higher XUV conditions, and it can decrease the escape rate to the critical flux against CO or $CO_2$ when the coolants' abundance and the cooling rate are enough to remove most of the radiative heating energy obtained by XUV absorption.

### 4.2 Dependence of the outflow structure on planetary mass

Although this study supposes an Earth-mass planet, the outflow structure such as molecular density profiles should depend on the planetary mass. In this section, we compare the results of this study with those of Yoshida and Kuramoto (2020), which investigated the hydrodynamic escape of Martian $H_2$-rich proto-atmospheres, to comprehend the effects of variations in planetary mass.

Compared with the outflow structure of Martian atmospheres, molecules including carbon oxides in an $H_2$-CO atmosphere on Earth-mass planet tend to decompose rapidly near the lower boundary (Figures 3(a)—(d); Figure 5 in Yoshida and Kuramoto, 2020 for an $H_2$-CO

Martian atmosphere). This is primarily because their photolysis proceeds more efficiently in the lower region under the slower advection in the deeper gravitational well. Instead of the decomposed molecules, chemical products such as $H_2O$ and $OH$ work as important coolants on heavier planets. The outflow temperature increases on heavier planets, as the thermal energy required for the occurrence of hydrodynamic escape is large due to their greater gravitational binding energy. The radiative cooling rate per molecule increases with temperatures as illustrated in Figure 1. Thus, the outflow tends to be more significantly suppressed by radiative cooling on heavier planets as long as molecular coolants are present in the outflow. The detailed investigation of the dependence on planetary mass is a focus of our future work.

### 4.3 Comparison with $H_2$-$CH_4$ atmospheres

This study focuses on the hydrodynamic escape of $H_2$-$CO$ and $H_2$-$CO_2$ atmospheres while Yoshida and Kuramoto (2021) investigated that of $H_2$-$CH_4$ atmospheres. In this section, we compare the outcomes of this study with those presented by Yoshida and Kuramoto (2021) to clarify the characteristics of the effects of carbon oxides on the hydrodynamic escape.

Figure 8 shows the atmospheric escape rates with the basal $C/H_2$ ratio. The escape rates of the $H_2$-$CO$ and $H_2$-$CO_2$ atmospheres start to decrease when the basal $C/H_2$ ratio exceeds ~$10^{-5}$ although that of the $H_2$-$CH_4$ atmospheres is almost constant while the basal $C/H_2$ ratio is lower than ~$10^{-3}$. The difference is mainly due to the difference in the abundance of radiatively active species in the region $r < \sim 2R_p$ where radiative heating mainly occurs. In the $H_2$-CO and $H_2$-$CO_2$ atmospheres, the main C-bearing species in the lower region with $r < \sim 2R_p$ is CO due to its relatively high immunity against photolysis. In addition, $H_2O$ and OH are also abundant in this region through the oxidation of $H_2$. On the other hand, the main C-bearing species in the same region of the $H_2$-$CH_4$ atmosphere is atomic carbon (Figure 1(c) in Yoshida and Kuramoto, 2021), which is radiatively inactive in the temperature range of the outflow, due to the efficient photolysis of $CH_4$ and high reactivity of the other C-bearing chemical products such as CH and $CH_3$. The existence of abundant coolants in the $H_2$-CO and $H_2$-$CO_2$ atmospheres leads to the more effective suppression of hydrodynamic escape rate even when the mixing ratio of C-bearing species is as low as $10^{-5}$. On the other hand, the escape rate of the $H_2$-$CH_4$ atmosphere becomes lower than that of the $H_2$-CO and $H_2$-$CO_2$ atmospheres when C/$H_2$ > ~0.005. This is due to the difference in the molecular mass of major C-bearing species. The escape rates of the $H_2$-CO and $H_2$-$CO_2$ atmospheres become regulated at the critical fluxes for CO and $CO_2$, respectively, in this range of the C/$H_2$ ratio. Meanwhile, the escape rate of the $H_2$-$CH_4$ atmosphere decreases continuously until it reaches the critical flux for $CH_4$, which is smaller than those for CO and $CO_2$ due to its smaller molecular mass (Equation (6)).

## 4.4 Evolution of early Earth's atmosphere

The suppressed hydrodynamic escape by the radiative cooling of carbon oxides and their photochemical products suggested by our calculation should lead to the prolongation of $H_2$-rich conditions and limitation of escape of heavier components such as C-bearing species. Here we examine the lifetime of $H_2$-rich proto-atmosphere containing CO or $CO_2$ as the primary C-bearing species based on the escape rate obtained from our calculation (Figure 5).

      We construct a time evolution model of atmospheric mass and composition with following assumptions. The lower atmosphere, that contains most of atmospheric mass, consists of $H_2$ and either of CO or $CO_2$. The initial amount of $H_2$ is given as a parameter and the initial amount of carbon oxides are determined so that the final amount of C-bearing species after the cessation of hydrodynamic escape reaches at 70 bar of equivalent $CO_2$, which approximates an estimated amount of carbon on the present-day Earth's surface layer (Holland, 1984). The XUV flux is 100 times higher than the present-day value, which is a feasible value for the time interval of interest (~ several 100 Myr) as previously mentioned. The radial distance of the homopause is fixed at $R_p$+1000 km. To highlight the role of hydrodynamic escape controlled by carbon oxides on the lifetime of $H_2$-rich atmosphere, we neglect other atmospheric escape processes, volatile delivery, photochemical conversion of CO or $CO_2$ in the lower atmosphere, and interaction between the atmosphere and the surface. Changes in atmospheric mass and

composition are numerically traced until the $H_2$ amount falls down to 0.1 bar, using the escape rate depending on atmospheric composition (Figure 5). Here the escape rates under the basal $CO/H_2$ or $CO_2/H_2$ ratio > 0.5, which are not obtained by our hydrodynamic simulation, are assumed to be the critical flux (Equation (6)) for $H_2$ and zero for carbon oxides. This assumption has little effect on the estimates of lifetime of $H_2$-rich atmosphere and the total loss of carbon oxides when initial $H_2$ amount > 10 bar equivalent since the period before the increase in the basal $CO/H_2$ or $CO_2/H_2$ to 0.5 largely occupies the lifetime.

The lifetime of $H_2$-rich atmosphere and the total loss amounts of CO and $CO_2$ depending on the initial $H_2$ amount are shown in Figure 9. For convenience, the atmospheric mass of each species is expressed by the equivalent surface pressure given by

$$P_i = \frac{M_i g}{4\pi R_p^2} \qquad (7)$$

where $P_i$ and $M_i$ are the equivalent surface pressure and atmospheric mass of species $i$, respectively. Here we define the lifetime for $H_2$-rich atmosphere as the time for the amount of $H_2$ to reach 0.1 bar equivalent. The lifetimes obtained for $H_2$-rich atmospheres containing CO or $CO_2$ are about one order of magnitude longer than that of the pure $H_2$ atmospheres due to the suppression of $H_2$ escape by radiative cooling and drag by carbon oxides (Figure 9(a)). The former exceeds 100 Myr when the initial $H_2$ amount is ~200 bar which is within estimates for early Earth (e.g., Yoshida and Kuramoto, 2021). The total fraction of lost CO and $CO_2$ are significantly low compared with that of $H_2$ (Figure 9(b)).

Although the atmospheric evolutionary calculations above do not consider $CH_4$, it can exist as a primary C-bearing species in early Earth's atmosphere due to the chemical reduction of surface volatiles by accreting metallic Fe (e.g., Kuramoto and Matsui, 1996; Schaefer and Fegley, 2010; Zahnle et al., 2020; Wogan et al., 2023). In $CH_4$-dominated phases, $H_2$ escape can be also suppressed to the critical flux against $CH_4$ by the radiative cooling effects of $CH_4$ and its photochemical products particularly when the $CH_4$ mixing ratio exceeds ~1% (Section 4.2; Yoshida and Kuramoto, 2021). This possibly further extends the lifetime of $H_2$-rich atmosphere by 2-3 times since the $H_2$ escape flux from $CH_4$-contaminated $H_2$-rich atmospheres are estimated to be 2-3 times lower. $CH_4$ is estimated to be gradually photolyzed to produce organics and carbon oxides (Zahnle et al., 2020; Wogan et al., 2023; Yoshida et al., 2024). Even if the $CH_4$ abundance is decreased by the photochemical processes, CO and $CO_2$ derived from photochemical oxidation of $CH_4$ as well as volcanic and impact degassing should continuously suppress $H_2$ escape as alternatives to $CH_4$.

Although this model neglects interactions with the surface, the mass and composition of $H_2$-rich atmospheres can be significantly influenced by chemical reactions with a magma ocean to produce $H_2O$ (e.g., Ikoma and Genda, 2006; Young et al., 2023). Recent studies on high-pressure silicate melting suggest that the disproportionation of ferrous iron ($Fe^{2+}$) to ferric iron ($Fe^{3+}$) and metallic iron (Fe), followed by the removal of metallic iron to the core and homogenization of the iron oxides through vigorous convection, would oxidize the magma ocean (e.g., Armstrong et al., 2019; Deng et al., 2020; Kuwahara et al., 2023), potentially oxidizing the overlying atmosphere as well. Future work will explore atmospheric evolution in response to chemical reactions with such an oxidized magma ocean.

Considering that the $H_2$ escape rate is regulated to the critical flux against major coolant chemical species, noble gases Kr and Xe likely suffer negligible hydrodynamic escape from early Earth's atmosphere containing C-bearing species because they have molecular masses larger than $CO_2$. Ne is marginally escapable in $H_2$-CO and $H_2$-$CO_2$ atmospheres and Ar is also in $H_2$-$CO_2$ atmospheres. But their expected depletion is moderate within a factor, not exceeding an order from their initial amounts applying the established mass fractionation theory. This result differs from the traditional picture for the escape of a pure solar nebular atmosphere, where the noble gases efficiently escape to space along with hydrogen (Sekiya et al., 1980; Sasaki and Nakazawa, 1988; Pepin, 1991).

On the other hand, proto-Earth is theoretically estimated to gravitationally capture the solar nebular gas as its proto-atmosphere during accretion, with the accumulated atmospheric mass depending on parameters such as proto-planetary mass and mass accretion rate in the solar nebula (e.g., Nakazawa et al., 1985; Ikoma and Genda, 2006; Lammer et al., 2014). The minimal escape of the noble gases from $H_2$-dominated atmosphere enriched in C-bearing species suggests that most of the noble gases derived from the solar nebula were lost before Earth's accretion was complete, to align with the depletion of the noble gases in the Earth's atmosphere compared with the solar composition. On proto-planets that were embryos of Earth with masses similar to that of Mars, the amount of the gravitationally captured solar nebular gas is estimated to have been limited compared to heavier planets (Ikoma and Genda, 2006; Lammer et al., 2014; Saito and Kuramoto, 2018). Furthermore, atmospheric loss via hydrodynamic escape and giant impact events should have occurred more efficiently on smaller-mass proto-planets (Genda and Abe, 2005; Odert et al., 2018; Yoshida and Kuramoto, 2020). These factors would explain the Earth's depletion of noble gases even if those Earth's embryos initially acquired nebula-derived atmospheres.

The hindered escape of heavy species, such as C- and N-bearing species, prevents isotopic fractionation of these elements. This can explain the similarity between the carbon and nitrogen isotopic compositions of Earth's surface inventory and primitive meteorites, which likely represent Earth's building blocks (Marty, 2012), as discussed in detail by Yoshida and Kuramoto (2021). On the other hand, Ne and Xe in Earth's atmosphere have been known enriched in heavier isotopes relative to the solar composition (e.g., Pepin, 1991, 2006; Marty, 2012). The enrichment of heavier isotope of Ne may be explained by mixing between the solar component and chondritic component (Marty, 2012) and/or hydrodynamic escape from low-mass proto-planets. Recent analysis of Xe isotopic compositions in Archean sedimentary rocks suggests that the Xe depletion and isotopic fractionation progressed over the first half of Earth's history (e.g., Pujol et al., 2011; Avice et al., 2018; Warr et al., 2018; Zahnle et al., 2019; Avice et al., 2020; Broadley et al., 2022; Ardoin et al., 2022), which implies that Xe isotopic fractionation was driven by the escape of Xe ions carried by hydrogen ions from a weakly reduced atmosphere during the Archean, rather than by early-stage hydrodynamic escape (Zahnle et al., 2019).

The prolongation of $H_2$-rich conditions may have contributed to the warming of early Earth under the faint young Sun by the greenhouse effect through line absorptions and collision-induced absorptions (e.g., Sagan and Mullen, 1972; Ramirez et al., 2014). Moreover, the sustained $H_2$-rich reduced environments may have played important roles in prebiotic chemical evolution since the synthesis of various organics and prebiotically important molecules such as $H_2CO$ and HCN, which are precursors to amino acids, nucleobases, and sugars, can proceed efficiently through reactions induced by the irradiation of high-energy photons and particles, lightning, and/or impact shock in the reduced environments (e.g., Schlesinger and Miller, 1983).

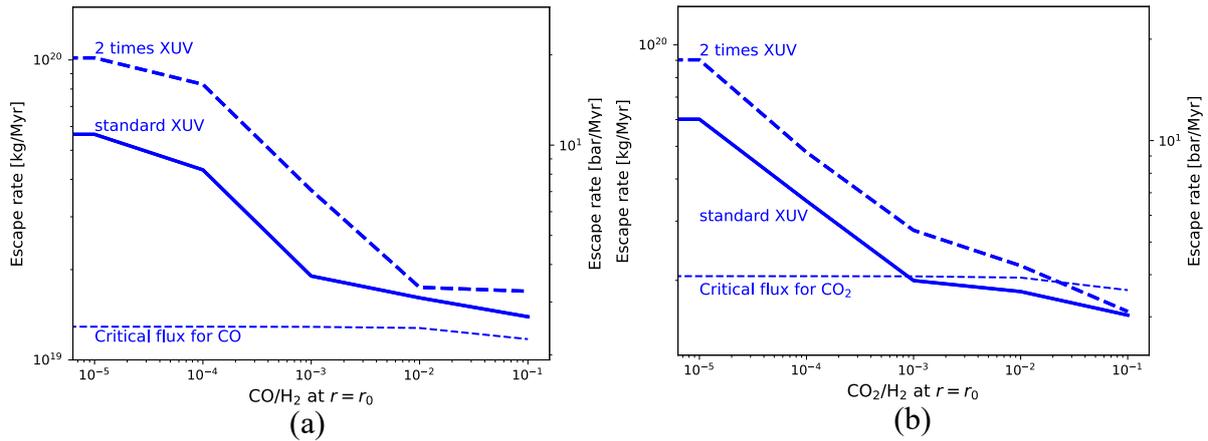

Figure 7. Comparison of the total atmospheric escape rates when the XUV flux is twice the standard setting with those under the standard setting on the $H_2$-CO atmospheres (a) and $H_2$-$CO_2$ atmospheres (b), respectively. The solid lines represent the values under the standard setting, and the dashed lines represent the values when the XUV flux is twice the standard setting. The right vertical axis represents the escape rate normalized by the mass of the present-day Earth's atmosphere. The thin dashed lines show the critical fluxes, which are the minimum fluxes of $H_2$ that can drag other heavier species outside the planetary gravitational well, as described by Equation (6).

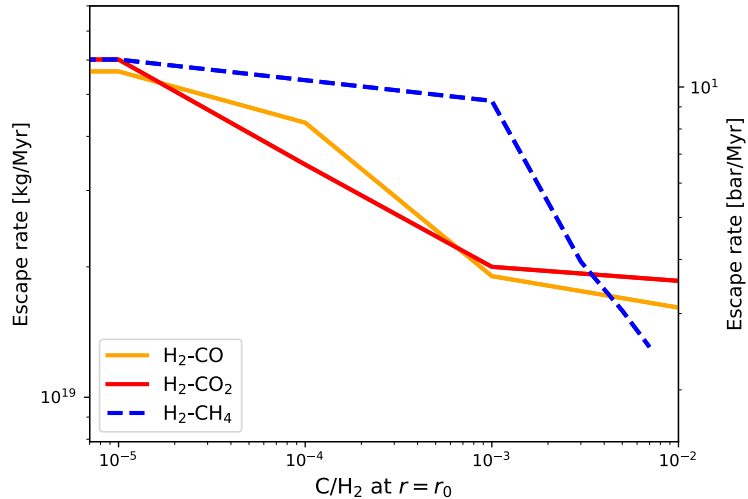

Figure 8. Total atmospheric escape rates of the $H_2$-CO atmospheres (orange), $H_2$-$CO_2$ atmospheres (red), and $H_2$-$CH_4$ atmospheres (dashed blue), respectively. The right vertical axis represents the escape rate normalized by the mass of the present-day Earth's atmosphere.

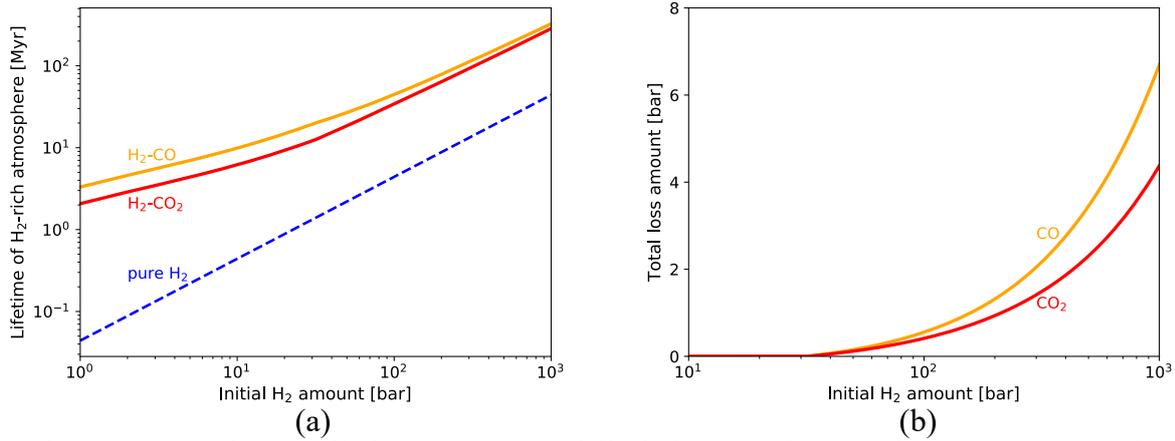

Figure 9. (a) Time until the H$_2$ amount falls below 0.1 bar on the H$_2$-CO atmospheres (orange), H$_2$-CO$_2$ atmospheres (red), and pure H$_2$ atmosphere (dashed blue). The lifetime for the pure H$_2$ atmosphere is obtained by dividing the initial H$_2$ amount by the escape rate of pure hydrogen atmosphere. (b) Total loss amounts of CO from H$_2$-CO atmospheres (orange) and CO$_2$ from H$_2$-CO$_2$ atmospheres (red) normalized by 1 bar until the H$_2$ amount falls below 0.1 bar.

## 5 Conclusions

We have conducted 1-D hydrodynamic escape simulations for H$_2$-rich atmospheres incorporating CO, CO$_2$, and their chemical products on an Earth-mass planet by considering detailed radiative cooling processes and chemical networks related to carbon oxides. According to our results, the atmospheric escape rate decreases as the basal CO/H$_2$ or CO$_2$/H$_2$ ratio increases due to the radiative cooling effects of CO, H$_2$O, OH, and H$_3^+$. Specifically, the H$_2$ escape rate from the H$_2$-CO (or H$_2$-CO$_2$) atmosphere asymptotes the critical flux for CO (or CO$_2$) when the basal CO/H$_2$ (or CO$_2$/H$_2$) ratio reaches ~0.001. These results suggest a prolongation of the timescale for H$_2$ escape in the presence of CO or CO$_2$, extending by about one order of magnitude compared to the case of pure hydrogen atmospheres, while also resulting in negligible escape of heavier species such as carbon- and nitrogen-bearing species, and noble gases.

## Abbreviations

XUV: X-ray and extreme ultraviolet

## Declarations

### Availability of data and material

Data sharing not applicable to this article as no datasets were generated or analysed during the current study. Please contact author for data requests.

### Competing interests

The authors declare that they have no competing interest.

### Funding

This work was supported by JSPS KAKENHI Grant Number 21K03638, 21H01155, 22K21344, 22H00164, 23KJ0093, and 23H04645.

**Authors' contributions**

TY proposed the topic, designed the study, and carried out the simulations. NT and KK helped in their interpretation and collaborated with the corresponding author in the construction of the manuscript. All authors read and approved the final manuscript.

298-306.

Ramirez, R. M., Kopparapu, R., Zugger, M. E., Robinson, T. D., Freedman, R., & Kasting, J. F. (2014). Warming early Mars with CO2 and H2. *Nature Geoscience*, *7*(1), 59-63.

Rothman, L. S., Gordon, I. E., Babikov, Y., Barbe, A., Benner, D. C., Bernath, P. F., et al. (2013). The HITRAN2012 molecular spectroscopic database. *Journal of Quantitative Spectroscopy and Radiative Transfer*, *130*, 4-50.

Sagan, C., & Mullen, G. (1972). Earth and Mars: Evolution of atmospheres and surface temperatures. *Science*, *177*(4043), 52-56.

Saito, H., & Kuramoto, K. (2018). Formation of a hybrid-type proto-atmosphere on Mars accreting in the solar nebula. *Monthly Notices of the Royal Astronomical Society*, *475*(1), 1274-1287.

Sasaki, S., & Nakazawa, K. (1988). Origin of isotopic fractionation of terrestrial Xe: hydrodynamic fractionation during escape of the primordial H2He atmosphere. *Earth and*
38

**Figure legends**

Figure 1. Radiative cooling rate per molecule under optically thin and LTE conditions as a function of temperature. The upper panel shows the cooling rates of C-bearing species, while the lower panel displays those of H- and O-bearing species, respectively. The net radiative cooling rate of each species is calculated by summing the cooling rate for each energy transition (Equation (12) in Supplementary Information), with reference to line data provided by HITRAN and ExoMol database.

Figure 2. Velocity and temperature profiles of the $H_2$-CO atmospheres (left column) and $H_2$-$CO_2$ atmospheres (right column). (a): Mean velocity of the $H_2$-CO atmospheres. The dots and triangles represent the transonic points and exobases, respectively. (b): Mean velocity of the $H_2$-$CO_2$ atmospheres. (c): Temperature of the $H_2$-CO atmospheres. (d): Temperature of the $H_2$-$CO_2$ atmospheres.

Figure 3. Number density and heating/cooling rate profiles of the $H_2$-CO atmospheres with the basal $CO/H_2=0.001$ (left column) and $H_2$-$CO_2$ atmospheres with the basal $CO_2/H_2=0.001$ (right column). (a): Number densities of selected neutral species in the $H_2$-CO atmosphere. (c): Number densities of selected ion species in the $H_2$-CO atmosphere. (e): Radiative heating rate (red solid line), radiative cooling rates (dashed lines), and chemical energy expense rate (purple dotted line) in the $H_2$-CO atmosphere, respectively. (b), (d), and (f) are the same as (a), (c), and (e), respectively, but for the $H_2$-$CO_2$ atmospheres.

Figure 4. (a) Rates of radiative heating, radiative cooling, and chemical energy expense (upper panels) and heating efficiency defined as Equation (4) (lower panels) in the subsonic region of the $H_2$-CO atmospheres as functions of the basal $CO/H_2$ ratio. (b) Same as (a), but for the $H_2$-$CO_2$ atmospheres.

Figure 5. (a) Escape rate of main species per 1 Myr as a function of the basal $CO/H_2$ ratio on the $H_2$-CO atmospheres. (b) Same as (a), but for the $H_2$-$CO_2$ atmospheres. The escape rates of CO (orange) and $CO_2$ (red) are multiplied by 100. The right vertical axis represents the mass flux normalized by the mass of the present-day Earth's atmosphere: 1 bar $=$ $5.3 \times 10^{18}$ kg. The dashed lines show the critical flux, which are the minimum fluxes of $H_2$ that can drag other heavier species outside the planetary gravitational well, as described by Equation (6).

Figure 6. Fractionation factor, defined as $(F_i/F_{H_2})/(n_i(r_0)/n_{H_2}(r_0))$ (where $i =$ CO, $CO_2$), as a function of the basal $CO/H_2$ or $CO_2/H_2$ ratio. Solid lines represent the numerical results, while dashed lines show the values derived from the analytical formula in Equation (8). The orange lines represent the fractionation factors between $H_2$ and CO in $H_2$-CO atmospheres, and the red lines represent those between $H_2$ and $CO_2$ in $H_2$-$CO_2$ atmospheres.

Figure 7. Comparison of the total atmospheric escape rates when the XUV flux is twice the standard setting with those under the standard setting on the $H_2$-CO atmospheres (a) and $H_2$-$CO_2$ atmospheres (b), respectively. The solid lines represent the values under the standard setting, and the dashed lines represent the values when the XUV flux is twice the standard setting. The right vertical axis represents the escape rate normalized by the mass of the present-day Earth's atmosphere. The thin dashed lines show the critical fluxes, which are the minimum fluxes of $H_2$ that can drag other heavier species outside the planetary gravitational well, as described by Equation (6).

Figure 8. Total atmospheric escape rates of the $H_2$-CO atmospheres (orange), $H_2$-$CO_2$ atmospheres (red), and $H_2$-$CH_4$ atmospheres (dashed blue), respectively. The right vertical axis

represents the escape rate normalized by the mass of the present-day Earth's atmosphere.

Figure 9. (a) Time until the $H_2$ amount falls below 0.1 bar on the $H_2$-CO atmospheres (orange), $H_2$-$CO_2$ atmospheres (red), and pure $H_2$ atmosphere (dashed blue). The lifetime for the pure $H_2$ atmosphere is obtained by dividing the initial $H_2$ amount by the escape rate of pure hydrogen atmosphere. (b) Total loss amounts of CO from $H_2$-CO atmospheres (orange) and $CO_2$ from $H_2$-$CO_2$ atmospheres (red) normalized by 1 bar until the $H_2$ amount falls below 0.1 bar.